# Exploring User Acceptance Of Portable Intelligent Personal Assistants: A Hybrid Approach Using PLS-SEM And fsQCA

*Gustave Florentin Nkoulou Mvondo\*[1], Ben Niu[1]*

**Abstract**

This research explores the factors driving user acceptance of Rabbit R1, a newly developed portable intelligent personal assistant (PIPA) that aims to redefine user interaction and control. The study extends the technology acceptance model (TAM) by incorporating artificial intelligence-specific factors (conversational intelligence, task intelligence, and perceived naturalness), user interface design factors (simplicity in information design and visual aesthetics), and user acceptance and loyalty. Using a purposive sampling method, we gathered data from 824 users in the US and analyzed the sample through partial least squares structural equation modeling (PLS-SEM) and fuzzy set qualitative comparative analysis (fsQCA). The findings reveal that all hypothesized relationships, including both direct and indirect effects, are supported. Additionally, fsQCA supports the PLS-SEM findings and identifies three configurations leading to high and low user acceptance. This research enriches the literature and provides valuable insights for system designers and marketers of PIPAs, guiding strategic decisions to foster widespread adoption and long-term engagement.

1. Introduction

Imagine having an intelligent personal assistant (IPA) that responds to commands and performs tasks across various apps or services without requiring users to navigate complicated interfaces or juggle multiple logins on smartphones or computers (Medium, 2024). This is the concept introduced by Rabbit R1, a newly developed portable IPA (PIPA) designed to revolutionize user interaction and control (Sullivan, 2024). Rabbit R1 aims to simplify traditional app-based interfaces by offering a single, unified interface for various tasks, such as controlling music, ordering services, and facilitating communication (The Verge, 2024). Its intuitive voice controls, combined with a dedicated training mode and adaptive learning capabilities, enable it to learn and execute tasks autonomously (Medium, 2024).

[1]College of Management, Shenzhen University, China (*Corresponding Author: florent90@yahoo.fr)

The recent advancements in artificial intelligence (AI) and natural language processing (NLP) software have led to the development of devices like Rabbit R1. The PIPA operates on a unique AI technology known as the large action model (LAM) (Sullivan, 2024). LAM extends beyond the generative outputs of large language models (LLMs) to include control over device applications (Zulhusni, 2024). The model's training involved numerous recorded user sessions with various apps, and its development was further enhanced through a partnership with the LLM-powered chatbot Perplexity AI, ensuring more precise and up-to-date responses to user inquiries (Cadenas, 2024). Rabbit R1 has garnered significant attention since its launch in January 2024, with its presentation receiving 11 million views at launch alone. The device sold out (10,000 units) in 24 hours, generating USD 10 million, and received 50,000 pre-orders within two weeks after its release (Brenner, 2024; Zulhusni, 2024).

Despite this enthusiasm, there is limited understanding of the factors motivating users' acceptance of PIPAs like Rabbit R1. Current research primarily focuses on digital IPAs such as Apple Siri (Brill et al., 2022), Amazon Alexa (Liao et al., 2019; McLean & Osei-Frimpong, 2019), and ChatGPT (Camilleri, 2024; Mvondo et al., 2024), leaving a crucial gap in our understanding of PIPAs. Unlike their digital counterparts, PIPAs combine AI capabilities with tangible hardware (Zulhusni, 2024), introducing unique dimensions of user interaction not addressed by research on digital assistants. This combination of physical form, portability, and AI functionality in PIPAs presents new challenges and opportunities for user engagement. Understanding the adoption factors for PIPAs is vital, as it will inform the design and development of more user-centric features, ensuring these devices meet user needs and preferences.

This research leverages the technology acceptance model (TAM) to explain user acceptance of Rabbit R1. TAM is a prominent model in information systems research that aims to explain people's intentions to adopt technology (Davis, 1989). The model posits that "perceived usefulness (PU) and perceived ease of use (PEOU) are key determinants of an individual's intention to use technology" (Niu & Mvondo, 2024, p. 3). TAM is widely employed due to its generic nature (e.g., Attié & Meyer-Waarden, 2022; Okine et al., 2023; Rodríguez-López et al., 2024; Türker et al., 2022). However, this ubiquity is also its primary limitation (Lew et al., 2020), as the model is inadequately equipped to fully explain portable AI companions like Rabbit R1. Hence, this study extends the model by incorporating critical aspects of AI and user interface design: perceived intelligence and simplicity.

Perceived intelligence refers to the extent to which users believe an AI system demonstrates cognitive capabilities, such as performing tasks efficiently, engaging in coherent and meaningful conversations, and exhibiting human-like qualities in its behavior and responses (Ling et al., 2023). It is an essential attribute of any AI system (Balakrishnan & Dwivedi, 2024), influencing how users evaluate and engage with technology (Guha et al., 2023; Moussawi et al., 2021; Priya & Sharma, 2023). Despite extensive research, the theoretical understanding of perceived intelligence remains unclear (Zhao et al., 2024). Much of the existing research has adopted a single-dimensional approach, neglecting the complexity of its various aspects and their impact on user acceptance and loyalty to IPAs (e.g., Balakrishnan & Dwivedi, 2024; Moussawi et al., 2020). The few studies that have employed a multidimensional approach have mainly examined booking and exploration intentions (Ling et al., 2023; Zhao et al., 2024) and did not explore how these aspects influence user acceptance and loyalty. Recent research suggests the need for a better understanding of the characteristics of perceived intelligence (Zhao et al., 2024). This understanding will allow for the identification of key aspects that are important to users and help companies develop products and services that specifically respond to users' needs and expectations.

Perceived simplicity is a critical element in user interface design, referring to features that facilitate the perception and processing of digital devices (Choi & Lee, 2012; Verkijika & De Wet, 2019). This concept encompasses facets such as information design and visual simplicity (Choi & Lee, 2012). Simplicity is essential for creating aesthetically pleasing and efficient interfaces, leading to positive evaluations of technological products (Choi & Lee, 2012). Despite its significance, it has not yet been explored in the context of IPAs. Additionally, its potential impact on user acceptance and loyalty remains underexplored in the literature. Examining its impact on user acceptance and loyalty to IPAs can offer valuable insights for designing more appealing and intuitive devices, ultimately enhancing user experience and driving long-term engagement.

This research aims to address the identified gaps in the literature on IPAs. It seeks to answer the question:

**RQ1: What key factors influence user acceptance of PIPAs like Rabbit R1?**

**RQ2: How does user acceptance drive loyalty?**

We develop an extended TAM model that integrates perceived intelligence and simplicity factors, along with user loyalty. To test our model, the study utilizes a sample of 821 users and employs advanced analytical methods such as partial least squares structural equation modeling (PLS-SEM) and fuzzy set qualitative comparative analysis (fsQCA).

The originality and contributions of this research are as follows. Theoretically, as a pioneering study examining user acceptance of PIPAs like Rabbit R1, this study contributes to the literature on IPAs and human-computer interaction (HCI). It reveals the positive influence of PU, PEOU, perceived intelligence factors, and simplicity factors on user acceptance, which translates into user loyalty. Notably, fsQCA analyses support the PLS-SEM findings and identify configurations leading to high and low user acceptance. From a managerial perspective, this research provides valuable insights for system designers and marketers of PIPAs, guiding strategic decisions to foster widespread adoption and long-term engagement.

## 2. Literature review on IPAs

Sun (2021) defined an IPA as a software agent or device that provides services and solutions to users. Users can interact with IPAs either through text (text-based personal agents) or voice commands (voice-based personal agents) (Canbek & Mutlu, 2016). While most IPAs are virtual (e.g., ChatGPT and Apple Siri), some take the form of physical robots, such as Samsung Ballie, or portable devices, exemplified by Rabbit R1.

Scholars have increasingly focused on understanding the determinants of user adoption of IPAs. They have examined different types of assistants, such as virtual companions, digital pets, virtual health counselors, and chatbots integrated into various applications (Ling et al., 2021). These agents were explored in various contexts, ranging from general assistance to companionship, social support, communication, customer service, education, and healthcare (e.g., Badghish et al., 2024; Bhutoria, 2022; Kim et al., 2021; Kautish et al., 2023; Santiago et al., 2024; Song et al., 2022). Most of these studies have used TAM (Davis, 1989) and its variations, including TAM2 and the unified theory of acceptance and use of technology (UTAUT and UTAUT2), as theoretical foundations (e.g., Aw et al., 2022; Balakrishnan & Dwivedi, 2024; Vimalkumar et al., 2021). Other studies have used theories such as the uses and gratifications theory (U&G), the computers as social actors (CASA) paradigm, and the consumer acceptance of technology model (CAT-Model) (Kopplin, 2022; Xie et al., 2024).

The predictors of IPA adoption can be categorized into four main areas: usage-related factors, agent-related factors, user-related factors, and attitude and evaluation factors. Factors in the usage-related category involve how the technology is deployed and how users perceive its performance in specific contexts (Ling et al., 2021). This is the most prominent category, with many factors derived from TAM and UTAUT (e.g., Kang et al., 2024; Kautish et al., 2023; Sun, 2021). In the agent-related category, research mainly focuses on factors related to the social behavior of IPAs, with elements such as anthropomorphism and social presence receiving significant attention (Ling et al., 2021; Priya & Sharma, 2023). Research on user-related factors examines demographics such as age, gender, household size, and various psychological factors reflecting users' emotional states (e.g., Kaya et al., 2024; Priya & Sharma, 2023). For instance, McLean and Osei-Frimpong (2019) revealed that age, gender, and technology familiarity did not influence the use of IPAs. However, they observed that household size positively influenced acceptance. Finally, regarding the attitude and evaluation category, research has employed factors related to user attitudes and emotional reactions toward IPAs (Ling et al., 2021).

## 2.1. Perceived intelligence

Intelligence is an important characteristic of any AI-powered device (Balakrishnan & Dwivedi, 2024; Ling et al., 2023; Zhao et al., 2024). Intelligent systems have played a significant role in information systems since the 1950s (Moussawi et al., 2021). These systems were initially created to solve complex problems humans cannot solve. With the rapid advancement in AI, systems capable of answering questions in natural languages and exhibiting social skills have emerged (Balakrishnan & Dwivedi, 2024; Moussawi et al., 2021). Perception of intelligence in a system "denotes a combination of intelligent quotient and capabilities" (p. 4). It is how well a system can understand and process natural language input to provide relevant and useful information efficiently (Balakrishnan & Dwivedi, 2024; Ling et al., 2023). Moussawi et al. (2021) explain that a system is perceived as intelligent when "it understands user requests, responds without continuous user intervention, exhibits awareness of the physical and virtual worlds, adapts to change, learns from acquired information and user behavior, and completes tasks within a favorable timeframe" (p. 5).

Scholars have explored the perceived intelligence of IPAs from various perspectives. Moussawi et al. (2021) discovered that perceived intelligence affects anthropomorphism, PEOU, and PU. Similarly, Balakrishnan and Dwivedi (2024) found that it influenced user attitude and purchase intention. Priya and Sharma (2023) identified a positive relationship between perceived intelligence and utilitarian attitude, while Guha et al. (2023) revealed that it leads to a positive evaluation of IPAs. However, these studies have generally treated perceived intelligence as a unidimensional construct and did not validate its role in driving user acceptance and loyalty to IPAs. Moreover, the scales used in these studies do not adequately capture the unique characteristics of IPAs (Zhao et al., 2024), whose perception of intelligence largely depends on the textual characteristics of the dialogue, the ability to perform tasks autonomously, and human-like traits (Ling et al., 2023). Moussawi and Koufaris (2021) argued that treating perceived intelligence as a single-dimensional construct lacks sufficient depth in accurately measuring it. They advocated for a comprehensive approach that captures all aspects of intelligent systems. Thus, this research conceptualizes perceived intelligence as a construct with three dimensions specifically reflecting the unique characteristics of PIPAs: conversational intelligence, task intelligence, and perceived naturalness.

Conversational intelligence is the ability of IPAs to comprehend and process users' natural language inputs smoothly and intuitively (Ling et al., 2023). This capability is achieved through "natural language understanding/processing, machine learning, conversation context recognition, and memory" (Ling et al., 2023). Conversationally intelligent systems enhance user interactions by interpreting user queries, predicting user needs, and providing relevant and timely responses (Guha et al., 2023; Ling et al., 2023; Moussawi et al., 2021). By leveraging these technologies, conversationally intelligent systems aim to make interactions feel natural and proactive, ensuring that users can communicate with the device as effortlessly as they would with another person.

Task intelligence refers to the ability of IPAs to understand specific objectives or requests issued by users and execute associated tasks or commands proficiently, effectively, and in a contextually relevant manner (Moussawi & Koufaris, 2019; Zhao et al., 2024). This involves understanding the user's instructions and integrating contextual awareness to ensure the tasks are carried out accurately and appropriately (Guha et al., 2023; Moussawi et al., 2021). Task intelligence enables PIPAs to perform tasks and implement actions with minimal user intervention, such as controlling apps or devices, searching for information, and providing reminders or alerts (Hu et al., 2021a; Moussawi et al., 2021).

Perceived naturalness refers to the extent to which users believe the voice of an IPA resembles that of a human. Users often attribute human-like characteristics to technologies (Waytz et al., 2010). Perceived naturalness focuses on the user's impression of the speech quality of IPAs (Guha et al., 2023; Kang et al., 2024); more natural-sounding speech can lead users to attribute human-like traits to the device (Humphry & Chesher, 2021; Kang et al., 2024). This perception is influenced by factors such as tone, inflection, pauses, and conversational flow. Guha et al. (2023) explain that the naturalness of speech affects whether users perceive an IPA as artificial or intelligent. When speech sounds more natural, users may believe they are conversing with a human rather than an AI device (Hu et al., 2021b; Kang et al., 2024). Perceived naturalness differs from conversational intelligence in that the latter relates to the system's technical ability to manage conversations effectively, while the former concerns the user's perception of the IPA's voice quality and how human-like it sounds.

These aspects of perceived intelligence were chosen because they address essential elements that determine how users perceive and interact with PIPAs like Rabbit R1. Conversational intelligence ensures that the PIPA can understand and respond to natural language inputs smoothly, facilitating effective communication. Task intelligence focuses on the PIPA's ability to accurately execute user commands and perform tasks, which is crucial for its practical utility. Perceived naturalness pertains to the human-like quality of the PIPA's voice, influencing users' comfort and emotional connection with the device (Kang et al., 2024).

2.2. Perceived simplicity

Simplicity is an essential element of user interface design and HCI, reflecting the different facets that facilitate the perception and processing of a system (Choi & Lee, 2012; Verkijika & De Wet, 2019). It is the main component of beauty and is valued positively because it inherently provides pleasure to users, making their interactions with technology more enjoyable (Choi & Lee, 2012). It is intrinsic in nature, meaning that an object can be perceived as simple without extensive cognitive effort or detailed reasoning about its utility. Choi et al. (2012) explained that perceived simplicity is understood through its different facets, including simplicity in information design and visual aesthetics.

Simplicity in information design pertains to the optimal structuring of an interface (Choi & Lee, 2012). It involves designing a system so that only essential steps are required to complete a task (Choi & Lee, 2012; Verkijika & De Wet, 2019). This approach focuses on organizing, structuring, and presenting information clearly, efficiently, and effectively, creating a layout and visual hierarchy that helps users find, understand, and interact with information easily. Choi and Lee (2012) state that "interfaces that do not adhere to the principle of simplicity often include unnecessary and complicated steps in task execution." Consequently, users may experience increased difficulty navigating such devices, leading to heightened frustration (Verkijika & De Wet, 2019).

Visual aesthetics generally pertain to a digital interface or device's overall visual appeal and attractiveness (Choi & Lee, 2012). It refers to the degree to which users perceive a system's design or appearance as aesthetically appealing. An aesthetically pleasing design can engage users, making their interaction with the device more enjoyable and satisfying (Luo et al., 2024; Pang et al., 2024; Verkijika & De Wet, 2019). Visual simplicity is subjective and can vary among users (Verkijika & De Wet, 2019). However, to ensure that most users find an interface visually appealing, it is crucial to prioritize "unity in diversity" in interface design (Choi et al., 2012).

Perceived simplicity is an under-researched construct in HCI literature. Scant research has examined its impact on user acceptance and loyalty with technology. To the best of the authors' knowledge, research has yet to examine the construct in the context of IPAs. This may be because the extant literature has primarily focused on digital IPAs.

## 3. Theoretical foundation

### 3.1. TAM

TAM, developed by Davis (1989), is a framework designed to explain and predict user acceptance and adoption of technology. PU and PEOU are central constructs in TAM, as they determine user intention to adopt a technology. PU refers to "the subjective belief that a particular technology enhances one's performance or productivity," while PEOU reflects "the degree to which users perceive using a particular technology to be free of effort" (Davis, 1989). TAM has been widely employed in various contexts due to its simplicity and robustness, providing valuable insights into how user perceptions influence their acceptance and use of new technologies (Attié & Meyer-Waarden, 2022; Okine et al., 2023; Rodríguez-López et al., 2024). In the context of IPAs, several studies have extended TAM with contextual factors (e.g., Balakrishnan & Dwivedi, 2024;

Niu & Mvondo, 2024). For instance, Balakrishnan and Dwivedi (2024) extended TAM with factors such as perceived intelligence, animacy, and anthropomorphism, finding that the model could explain 33.3% and 38.9% of the variance in purchase intention and attitude toward IPAs. Likewise, Maduku et al. (2023) extended the model with privacy concern, localization, passion, commitment, and Word-of-mouth (WOM), finding that the model could explain 77.1%, 61.1%, and 96.3% of the variance in passion, WOM, and commitment, respectively.

## 4. Hypotheses development and research framework

### 4.1. PU and PEOU

PU and PEOU are key determinants of an individual's acceptance of technologies (Okine et al., 2023; Rodriguez-López et al., 2024; Türker et al., 2022), such as the Rabbit R1. Moussawi et al. (2021) explain that IPA characteristics, such as task completion and personalization, can influence users to perceive the device as useful. When users believe that Rabbit R1 can effectively streamline their activities—such as controlling various device applications, managing music, ordering services, and facilitating communication (The Verge, 2024)—they are more likely to acknowledge its value and be willing to integrate it into their daily lives. Additionally, if the device is designed with an interface that is user-friendly and requires a minimal learning curve (Balakrishnan & Dwivedi, 2024; Moussawi et al., 2021), users are more inclined to adopt it, as the simplicity reduces potential frustrations and cognitive load. Several scholars have established the relationship between PU, PEOU, and user intention to adopt IPAs (Balakrishnan & Dwivedi, 2024; Choi & Kim, 2023; Sohn & Kwon, 2020). In line with their findings, we propose the following hypotheses:

**H1.** PU is positively related to PIPA acceptance.

**H2.** PEOU is positively related to PIPA acceptance.

### 4.2. Perceived intelligence

#### 4.2.1. Conversational intelligence

As explained earlier, conversational intelligence is "the degree to which IPAs can comprehend users' natural language inputs and smoothly conduct conversations in a proactive and timely manner" (Ling et al., 2023, p. 5). While the extant literature has yet to establish the impact of conversational intelligence on PU and user acceptance of IPAs, Ling et al. (2023) argue that it is a critical element that can significantly influence the overall user interaction experience. When PIPAs like Rabbit R1 accurately understand and respond to users' spoken commands, interactions become more intuitive and efficient. Smooth, intuitive, and proactive conversations make users more likely to find the PIPA useful and valuable (Balakrishnan & Dwivedi, 2024; Priya & Sharma, 2023), leading to higher acceptance. This natural and responsive interaction fosters trust and reliability (Moussawi et al., 2021), influencing users to integrate the PIPA into their daily routines. Based on this reasoning, we hypothesize that:

**H3a.** Conversational intelligence is positively related to PU.

**H3b.** Conversational intelligence is positively related to PIPA acceptance.

### 4.2.2. Task intelligence

Task intelligence refers to the ability of IPAs to understand specific objectives or requests issued by users and execute associated tasks or commands proficiently, effectively, and in a contextually relevant manner (Moussawi & Koufaris, 2019; Zhao et al., 2024). While existing studies have not directly examined the impact of task intelligence on PU and IPA acceptance, we argue that when Rabbit R1 can effectively and accurately conduct tasks across various device applications, it can significantly influence users' perceptions of usefulness and their willingness to integrate the PIPA into their daily routines. This proficiency can reduce the user's effort and time to complete these tasks manually (Guha et al., 2023; Hu et al., 2021a), making the device a valuable tool for simplifying daily routines. However, if the PIPA struggles with task execution, it may lead to a perception of uselessness and hinder adoption. Several scholars have argued that the utility of IPAs can be higher because they can be used for various tasks (Balakrishnan & Dwivedi, 2024; Hu et al., 2021a). The ability to conduct tasks can influence user intention to adopt the AI companion (Moussawi et al., 2021). Based on this reasoning, we hypothesize that:

**H4a.** Task intelligence is positively related to PU.

**H4b.** Task intelligence is positively related to PIPA acceptance.

### 4.2.3. Perceived naturalness

Companies often design IPAs with natural human voices (Hu et al., 2021b; Kang et al., 2024; Priya & Sharma, 2023; Zhao et al., 2024). When PIPAs like Rabbit R1 utilize natural-sounding speech, users are more likely to perceive the AI companions as human-like. Features such as appropriate pauses, intonation, and conversational flow make the speech sound more authentic, reducing the cognitive effort required to understand and engage with the assistant (Guha et al., 2023; Humphry & Chesher, 2021). This naturalness of speech can foster a sense of familiarity and trust, making interactions more pleasant and less robotic (Hu et al., 2021b; Kang et al., 2024), thereby improving the overall experience with PIPAs. The extant literature has yet to establish the impact of perceived naturalness on IPA acceptance. However, Guha et al. (2023) explain that natural-sounding speech may lead users to conclude that they are talking to a human, which can influence user attitudes toward AI devices. Thus, we propose the following hypothesis:

**H5.** Perceived naturalness is positively related to PIPA acceptance.

### 4.3. Perceived simplicity

### 4.3.1. Simplicity in information design

Simplicity in information design is crucial for the success of a system (Choi et al., 2012). When a PIPA interface adheres to the principle of simplicity by eliminating unnecessary and complicated steps, it reduces cognitive load and potential frustration (Choi et al., 2012; Verkijika & De Wet, 2019). This straightforward interaction experience enhances users' perception of the device as user-friendly. As users find it simpler to accomplish their tasks, it can lead to positive experiences (Sharma et al., 2015; Verkijika & De Wet, 2019), making them more likely to adopt and integrate the AI companion into their daily routines. However, if the device is designed to be complex or includes unnecessary steps, users may find it difficult to navigate (Choi et al., 2012), hindering their acceptance. Consequently, simplicity in information design is crucial for ensuring that a large number of users adopt the device. Research has yet to thoroughly examine the impact of simplicity in information design on PEOU and IPA acceptance. However, Choi et al. (2012)

found that simplicity in design can lead to user satisfaction. Moreover, Sullivan (2024) explained that Rabbit R1 features a simple and minimalistic interface design, which may influence users' perception of user-friendliness and their acceptance of the device. Thus, we propose the following hypotheses:

**H6a.** Simplicity in information design is positively related to PEOU.

**H6b.** Simplicity in information design is positively related to PIPA acceptance.

### 4.3.2. Simplicity in visual aesthetics

This research argues that the simplicity in visual aesthetics can influence user acceptance. This is because a simple visual design can convey a sense of modernity and sophistication, making the device more attractive to users. Additionally, a product with an aesthetically well-designed interface is better valued (Luo et al., 2024; Pang et al., 2024; Verkijika & De Wet, 2019). By fostering a pleasant and straightforward user experience, simplicity in visual aesthetics can increase trust and willingness to adopt and regularly use Rabbit R1, thus promoting overall acceptance. Previous research has consistently shown that visual aesthetics positively influence users' intentions (Bölen, 2020; Kumar et al., 2021). Accordingly, we advance that:

**H7.** Visual aesthetics is positively related to PIPA acceptance.

### 4.4. User loyalty

Jones and Sasser (1995) described loyalty as "a feeling of attachment or affection for a company's people, products, or services" (p. 94). It involves consistently choosing one company's products and services over those of competitors (Chen et al., 2023; Maroufkhani et al., 2022). Loyal customers are less influenced by price or availability and are often willing to pay more to maintain the quality and familiarity they value. Loyalty is vital for a product's success and is reflected through ongoing usage and positive WOM (Niu & Mvondo, 2024). Continuance usage involves a user's repeated engagement and preference for the product, indicating a high level of satisfaction and value derived from it (Cui et al., 2022; Niu & Mvondo, 2024; Sohaib & Han, 2023; Zou et al., 2023). Likewise, positive WOM occurs when satisfied customers share their favorable experiences, endorsing the product and influencing potential buyers (Niu & Mvondo, 2024; Sohaib & Han, 2023). It is a unique method of spreading information based on personal recommendations rather than advertising or promotional efforts. This combination reinforces the customer's allegiance and helps to attract new customers, contributing to the product's success and the company's reputation.

This research argues that user acceptance of Rabbit R1 can lead to loyalty. When users find the PIPA effective and satisfying due to its utility, ease of use, intelligence, and simplicity in user interface design and visual aesthetics, they are more likely to integrate the device into their daily routines, resulting in consistent and long-term usage. Furthermore, a positive user experience enhances trust and satisfaction (Cui et al., 2022; Niu & Mvondo, 2024; Sohaib & Han, 2023; Zou et al., 2023), prompting users to recommend the device to others. The connection between user acceptance and loyalty has been examined in previous research. Alkhwaldi et al. (2022) revealed that users who adopt technological products or services often become loyal users. In line with this finding, we predict that user acceptance of PIPAs will lead to loyalty and mediate the relationships between the proposed factors and loyalty.

**H8.** PIPA acceptance is positively related to user loyalty.

**H9-H15.** PIPA acceptance positively mediates the association between the predictor variables (PU, PEOU, conversational intelligence, task intelligence, perceived naturalness, information design, and visual aesthetics) and user loyalty (Fig.1).

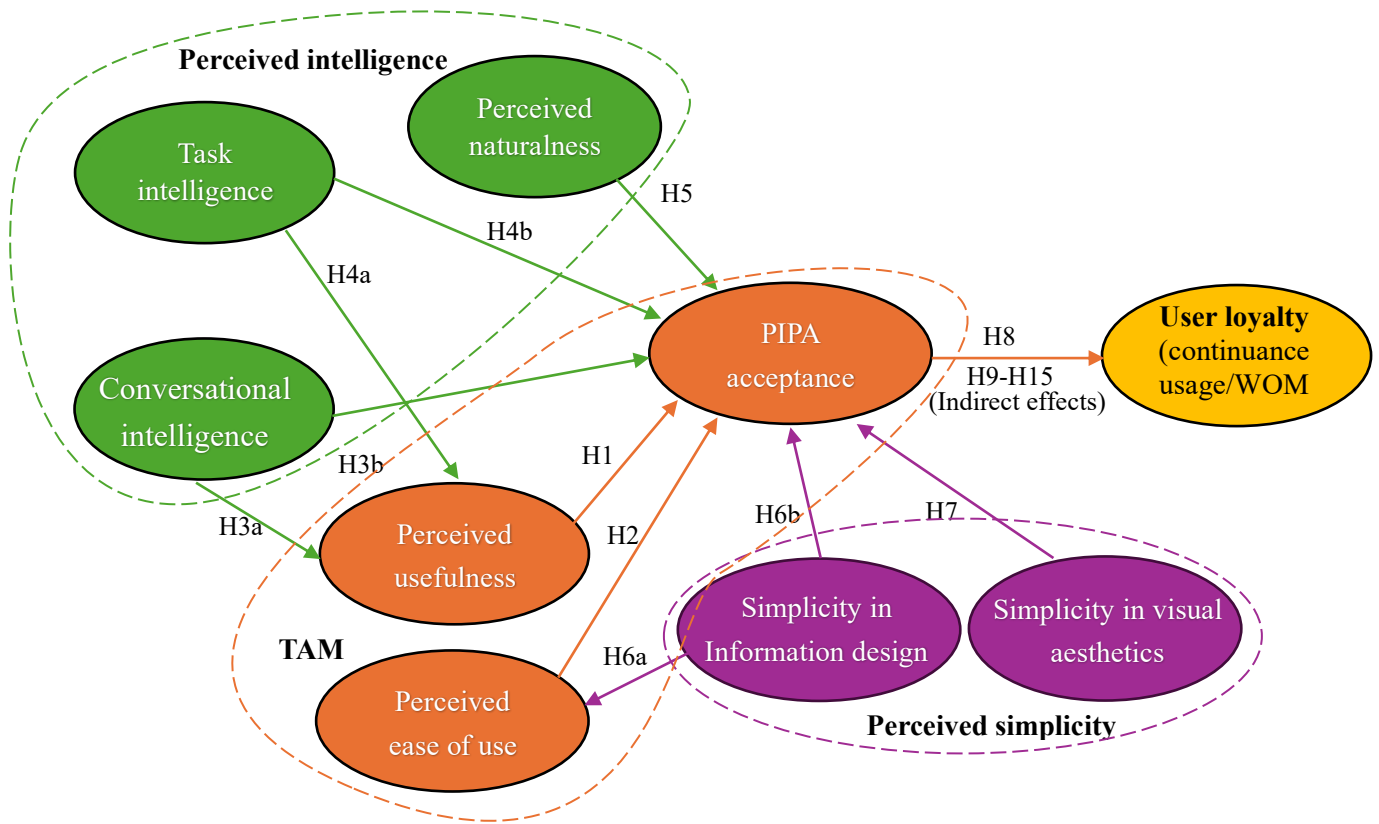

**Fig 1.** Theoretical framework

## 5. Methodology

### 5.1. Measures

We referred to the existing literature to develop the research instrument and items, making minor adjustments for contextual relevance. A 7-point Likert-type scale was used to measure the items, ranging from "1 (strongly disagree) to 7 (strongly agree)." The measurement items and their sources are detailed in Table 1.

**Table 1** Research instruments

| Instruments and Items | Sources |
|---|---|
| PU (3 items) and PEOU (4 items) | Davis (1989) |
| Conversational intelligence (4 items) | Ling et al. (2023) |
| Perceived naturalness (3 items) | Guha et al. (2023) and Hu et al. (2021b) |
| Task intelligence (3 items) | Moussawi & Koufaris (2019) and Hu et al. (2021a) |
| Information design (3 items) and visual aesthetics (3 items) | Choi & Lee (2012) |
| Acceptance (3 items) | Gursoy et al. (2019) |
| User loyalty (2 items) | Niu & Mvondo (2024) |

### 5.2. Pilot Study

We first carried out a small-scale pilot study with three primary objectives: evaluating the validity of the measurement scales, (2) enhancing the quality of the questions, and (3) refining and adjusting the survey. The pilot study involved 90 respondents (37 females). Based on respondents' feedback, exploratory factor analysis, and reliability analysis, we made further revisions and finalized the questionnaire, removing factors with loadings lower than 0.5.

### 5.3. Data collection and sampling

Data were collected from US residents in June 2024 using a structured online questionnaire administered through MTurk, employing a purposive sampling method. We specifically collected data in the US because Rabbit R1 was first made available there. To ensure the quality of our data and address threats posed by self-misrepresentation, inconsistent English language fluency, and MTurker non-naivete, we undertook several measures following guidelines outlined by Aguinis et al. (2021). First, we designed a concise questionnaire and avoided scales with labeled 'end' points to enhance the quality of the data. Second, to address inconsistent English language fluency, we determined *a priori* that only respondents from the US, a native English-speaking country, would be considered. Third, to address MTurker non-naivety, inattention, and vulnerability to web robots, we incorporated an attention check question. This open-ended question asked respondents what the study was about (Aguinis et al., 2021). Lastly, respondents who passed the attention check and took between four to ten minutes to complete the survey were retained, aligning with screening practices to detect careless responses.

This meticulous process resulted in 824 valid responses, with 489 males (59.3%) and 335 females (40.7%). Among the respondents, 43.8% identified as White, and 26.9% as Black or African American. The majority were aged 18-40 years. Additionally, 41.1% held a bachelor's degree, with most respondents being students (33.4%), followed by employees (32.2%) and businesspeople (24.9%) (see Table 2).

**Table 2** Demographic characteristics

| Items | Frequency (N=824) | (%) |
|---|---|---|
| ***Gender*** | | |
| Male | 489 | 59.3 |
| Female | 335 | 40.7 |
| ***Ethnicity*** | | |
| White | 361 | 43.8 |
| Black or African American | 222 | 26.9 |
| Asian | 132 | 16 |
| Hispanic | 88 | 10.7 |
| Other | 21 | 2.5 |
| ***Age*** | | |
| 18-30 Years | 299 | 36.3 |
| 31-40 Years | 305 | 37 |
| 41-50 Years | 174 | 21.1 |
| Above 51 Years | 46 | 5.6 |
| *Education* | | |
| High School or below | 151 | 18.3 |
| Bachelor's | 341 | 41.4 |
| Masters | 247 | 30 |
| Doctor (PhD) | 54 | 6.6 |
| Other | 31 | 3.8 |
| *Profession* | | |
| Student | 275 | 33.4 |
| Employee | 265 | 32.2 |
| Businesspeople | 205 | 24.9 |
| Unemployed | 42 | 5.1 |
| Other (e.g., Retired) | 37 | 4.5 |

5.4. Analytical method

This research employed a synergistic approach, combining PLS-SEM and fsQCA for data analysis. This strategy has gained increasing popularity among management scientists aiming to address the inherent limitations of each method (Acquah et al., 2024; Kang et al., 2024; Mvondo & Niu, 2024; Navarro-García et al., 2024), thereby enhancing the robustness of the research findings. PLS-SEM was chosen for several reasons: it is suitable for constructing and validating exploratory theoretical models. Since our study is exploratory and seeks to identify the drivers of PIPA acceptance through theoretical integration, using PLS-SEM is highly appropriate. Additionally, it has strong statistical power for hypothesis testing (Rasoolimanesh et al., 2021). Unlike conventional covariance-based methods, PLS-SEM uses bootstrapping techniques to ensure parameter stability and significance. Finally, as Hair et al. (2019) explain, PLS-SEM is particularly suitable when "the structural model is complex and includes many constructs, indicators, and/or model relationships" (p. 4). Given our model's complexity with seven independent variables, PLS-SEM is deemed appropriate.

However, PLS-SEM is limited in that it cannot identify the critical combinations of factors that lead to high or low acceptance. In contrast, fsQCA excels in identifying these combinations

(Fiss, 2011). fsQCA stands out due to its ability to handle intricate relationships among multiple predictors and outcome variables (Pappas et al., 2016). Unlike traditional symmetric analyses like regression and correlation, fsQCA considers multiple factors and their combinations, providing a nuanced understanding of causality through fuzzy set theory and Boolean algebra (Ragin, 2009). This approach addresses limitations often encountered in correlation-based methods, such as overfitting, and offers a more comprehensive view of complex relationships. Rasoolimanesh et al. (2021) suggest that combining PLS-SEM and fsQCA can offer deeper insights for theory and practice. Thus, we combined these two methods to examine the drivers and configuration paths of user acceptance of PIPAs. This approach will offer more effective managerial recommendations by considering various combinations of causal factors in specific organizational contexts.

The data analysis unfolded in three stages: (1) assessing model reliability and validity, (2) testing hypothesized relationships, and (3) conducting fsQCA. We employed SmartPLS version 4.1 for common method bias (CMB) and PLS-SEM, while for fsQCA, we utilized version 4.1.

## 6. Results

### 6.1. Common method bias

We employed a multistep approach to address CMB. Initially, we assured respondents of the study's strict anonymity protocol, emphasizing that their survey responses were intended to reflect their genuine experiences with Rabbit R1. This approach aimed to ensure unbiased responses free from external influences. Secondly, we conducted Harman's single-factor test, which indicated that a single factor explained only 9.599% of the total variance, well below the commonly accepted threshold of 50%. Lastly, following Kock's (2015) recommendation, we conducted a comprehensive collinearity test and confirmed that all variance inflation factors were below the threshold of 3.3. Therefore, we confidently assert that CMB did not significantly affect our study.

### 6.2. Measurement model

We assessed our measurement model's reliability, convergent, and discriminant validity. The findings indicated that the latent variables exhibited strong reliability, with factor loadings ranging from 0.856 to 0.762, Cronbach's alpha values between 0.877 and 0.780, and composite reliability (CR) consistently exceeding 0.7. Furthermore, the constructs' average variance extracted (AVE) ranged from 0.663 to 0.559, surpassing the critical threshold of 0.5. Our discriminant validity was substantiated through the Fornell and Larcker methods, where the square root of AVE exceeded correlations. The Heterotrait-Monotrait (HTMT) criterion also yielded values below 0.85, indicating clear differentiation between constructs. Moreover, we assessed multicollinearity using the variance inflation factor (VIF), and all VIF values remained below 5, confirming the absence of problematic multicollinearity (see Tables 3 and 4).

**Table 3** Reliability and validity

| Constructs | Items | Loading | Alpha >0.7 | CR >0.7 | AVE >0.5 |
|---|---|---|---|---|---|
| Perceived usefulness | PU1: Using Rabbit R enables me to accomplish tasks more quickly. | 0.849*** | 0.844 | 0.845 | 0.645 |
| | PU2: Using Rabbit R1 increases my productivity. | 0.799*** | | | |
| | PU3: I find Rabbit R1 very useful. | 0.759*** | | | |
| Perceived ease of use | PEU1: Learning to operate Rabbit R1 is easy for me. | 0.762*** | 0.835 | 0.835 | 0.559 |
| | PEU2: I find it easy to get Rabbit R1 to do what I want to do. | 0.692*** | | | |
| | PEU3: My interaction with Rabbit R1 is clear and understandable. | 0.765*** | | | |
| | PEU4: I find Rabbit R1 easy to use. | 0.768*** | | | |
| Conversational intelligence | CI1: Rabbit R1 understands the user very well. | 0.763*** | 0.877 | 0.877 | 0.641 |
| | CI2: Rabbit R1 remembers what the user asked previously. | 0.835*** | | | |
| | CI3: The conversation with Rabbit R1 flows smoothly. | 0.795*** | | | |
| | CI4: The conversation with Rabbit R1 flows naturally. | 0.807*** | | | |
| Task intelligence | TI1: Rabbit R1 can complete tasks quickly. | 0.825*** | 0.848 | 0.848 | 0.650 |
| | TI2: Rabbit R1 can find and process the necessary information for completing the tasks. | 0.801*** | | | |
| | TI3: Rabbit R1 can provide users with useful answers. | 0.793*** | | | |
| Perceived naturalness | PN1: Rabbit R1 sounds natural. | 0.795*** | 0.847 | 0.847 | 0.649 |
| | PN2: Rabbit R1 has a human-like voice. | 0.822*** | | | |
| | PN3: I cannot feel the distance between the voice of Rabbit R1 and that of a human being. | 0.800*** | | | |
| Simplicity in information design | SID1: Rabbit R1 has the necessary steps to operate. | 0.757*** | 0.834 | 0.834 | 0.626 |
| | SID2: Rabbit R1 has simple steps to use functions. | 0.849*** | | | |
| | SID3: Rabbit R1 has necessary functions. | 0.765*** | | | |
| Simplicity in visual aesthetics | SVA1: Rabbit R1 has a neat appearance. | 0.847*** | 0.854 | 0.855 | 0.663 |
| | SVA2: Rabbit R1 is modern. | 0.834*** | | | |
| | SVA3: Rabbit R1 interface design is well balanced. | 0.758*** | | | |
| Acceptance | ACC1: I am willing to use Rabbit R1. | 0.758*** | 0.841 | 0.841 | 0.638 |
| | ACC2: I feel happy to interact with Rabbit R1. | 0.803*** | | | |
| | ACC3: I am likely to interact with Rabbit R1. | 0.834*** | | | |
| User loyalty | UL1: I intend to continue using Rabbit R1 in the future and will keep using it regularly as I do now. | 0.746*** | 0.780 | 0.784 | 0.645 |
| | UL2: I will strongly recommend others to use Rabbit R1. | 0.856*** | | | |

**Table 4** Discriminant validity analysis

| Constructs | 1 | 2 | 3 | 4 | 5 | 6 | 7 | 8 | 9 |
|---|---|---|---|---|---|---|---|---|---|
| 1. ACC | **0.799** | | | | | | | | |
| 2. CI | 0.530 | **0.800** | | | | | | | |
| 3. PEOU | 0.294 | 0.152 | **0.748** | | | | | | |
| 4. PN | 0.539 | 0.608 | 0.187 | **0.806** | | | | | |
| 5. PU | 0.547 | 0.630 | 0.229 | 0.623 | **0.803** | | | | |
| 6. SID | 0.545 | 0.446 | 0.287 | 0.444 | 0.451 | **0.791** | | | |
| 7. SVA | 0.530 | 0.625 | 0.165 | 0.612 | 0.654 | 0.458 | **0.814** | | |
| 8. TI | 0.540 | 0.617 | 0.153 | 0.629 | 0.623 | 0.497 | 0.590 | **0.806** | |
| 9. UL | 0.390 | 0.231 | 0.338 | 0.203 | 0.218 | 0.238 | 0.217 | 0.213 | **0.803** |

6.3. Structural model

We first evaluated the predictive relevance of the model using the R-squared ($R^2$) approach. $R^2$ is "the proportion of an endogenous construct's variance explained by its predictor constructs in a regression model" (Mvondo et al., 2022a, 2022b). The analysis revealed that the extended TAM explains 48% of the variance in user acceptance.

We tested our hypothesized relationships by applying the statistical bootstrap technique with a sample size of 5,000. Our findings support all hypothesized relationships, including both direct and indirect effects. Specifically, we found that PU (β=0.117) and PEOU (β=0.120) were positively related to PIPA acceptance, validating H1 and H2. Additionally, conversational intelligence positively impacted PU (β=0.397) and PIPA acceptance (β=0.114), validating H3a and H3b.

Moreover, task intelligence positively influenced PU (β=0.378) and PIPA acceptance (β=0.114), supporting H4a and H4b. Furthermore, perceived naturalness was positively related to PIPA acceptance (β=0.131), validating H5. Simplicity in information design positively impacted PEOU (β=0.287) and PIPA acceptance (β=0.245), while visual aesthetics were positively related to PIPA acceptance (β=0.103), supporting H6 (a and b) and H7. Finally, PIPA acceptance was related to user loyalty (β=0.390), supporting H8.

Our findings also revealed that PIPA acceptance positively mediates the association between PU (β=0.046), PEOU (β=0.047), conversational intelligence (β=0.045), task intelligence (β=0.044), perceived naturalness (β=0.051), simplicity in information design (β=0.096), and visual aesthetics (β=0.040) and user loyalty (see Tables 5 and 6 for details).

**Table 5** Direct effect

| | Hypothesized relationships | Std. Beta | Std. Error | Supported |
|---|---|---|---|---|
| H1 | PU ➔ ACC | 0.117 | 0.051 | Yes |
| H2 | PEOU ➔ ACC | 0.120 | 0.030 | Yes |
| H3a | CI ➔ PU | 0.397 | 0.059 | Yes |
| H3b | CI ➔ ACC | 0.114 | 0.048 | Yes |
| H4a | TI ➔ PU | 0.378 | 0.059 | Yes |
| H4b | TI ➔ ACC | 0.114 | 0.051 | Yes |
| H5 | PN ➔ ACC | 0.131 | 0.056 | Yes |
| H6a | SID ➔ PEOU | 0.287 | 0.051 | Yes |
| H6b | SID ➔ ACC | 0.245 | 0.054 | Yes |
| H7 | SVA ➔ UA | 0.103 | 0.044 | Yes |
| H8 | ACC ➔ UL | 0.390 | 0.052 | Yes |

**Table 6** Indirect effects

| | Hypothesized relationships | Std. Beta | Std. Error | Supported |
|---|---|---|---|---|
| H9 | PU ➔ ACC ➔ UL | 0.046 | 0.021 | Yes |
| H10 | PEOU ➔ ACC ➔ UL | 0.047 | 0.015 | Yes |
| H11 | CI ➔ ACC ➔ UL | 0.045 | 0.020 | Yes |
| H12 | TI ➔ ACC ➔ UL | 0.044 | 0.021 | Yes |
| H13 | PN ➔ ACC ➔ UL | 0.051 | 0.023 | Yes |
| H14 | SID ➔ ACC ➔ UL | 0.096 | 0.024 | Yes |
| H15 | SVA ➔ ACC ➔ UL | 0.040 | 0.019 | Yes |

6.4. fsQCA analysis

### 6.4.1. Calibration

Calibration is a prerequisite step before performing fsQCA analysis. The data from PLS-SEM was calibrated into fuzzy sets. Fuzzy sets are on "a continuous scale ranging from 0 to 1, where 0 represents full non-set membership, and 1 signifies full set membership." As the survey employed a 7-point Likert scale, we set the full membership threshold at 7 and non-full membership at 1 (Pappas et al., 2016). Due to the non-normal distribution of our data, we determined the crossover point as the mean value of each condition, following the approach outlined by Fiss (2011). We utilized fsQCA 4.1 to transform variables into calibrated sets (see Table 8).

**Table 8** Calibration

|      | Full non-membership | Crossover point | Full membership |
|------|---------------------|-----------------|-----------------|
| PU   | 1                   | 4.94            | 7               |
| PEOU | 1                   | 5.32            | 7               |
| CI   | 1                   | 4.89            | 7               |
| TI   | 1                   | 4.92            | 7               |
| PN   | 1                   | 4.86            | 7               |
| SID  | 1                   | 5.00            | 7               |
| SVA  | 1                   | 4.89            | 7               |
| ACC  | 1                   | 5.01            | 7               |

### 6.4.2. Analysis of necessary conditions

The examination of necessary conditions aimed to identify if there are any specific causal conditions necessary for user acceptance of the IPA. Following Ragin's (2009) criteria, a condition is necessary when its consistency exceeds 0.9. The findings in Table 9 show that no single condition qualifies as necessary for achieving user acceptance (ACC). Additionally, no single condition is deemed necessary to negate user acceptance (~ACC). These results suggest that no individual conditions can independently lead to either ACC or ~ACC outcomes.

**Table 9** Analysis of necessary conditions

|            | Outcome variable: ACC | | Outcome variable: ~ACC | | |
|------------|-----------------------|----------|------------|-------------|----------|
| Conditions | Consistency | Coverage | Conditions | Consistency | Coverage |
| PU | 0.832 | 0.836 | PU | 0.733 | 0.522 |
| ~PU | 0.524 | 0.735 | ~PU | 0.769 | 0.764 |
| PEOU | 0.806 | 0.797 | PEOU | 0.810 | 0.567 |
| ~PEOU | 0.563 | 0.807 | ~PEOU | 0.710 | 0.721 |
| CI | 0.837 | 0.840 | CI | 0.738 | 0.525 |
| ~CI | 0.527 | 0.740 | ~CI | 0.776 | 0.771 |
| TI | 0.831 | 0.832 | TI | 0.734 | 0.520 |
| ~TI | 0.521 | 0.734 | ~TI | 0.763 | 0.761 |
| PN | 0.826 | 0.834 | PN | 0.730 | 0.522 |
| ~PN | 0.526 | 0.733 | ~PN | 0.768 | 0.758 |
| SID | 0.834 | 0.839 | SID | 0.738 | 0.526 |
| ~SID | 0.529 | 0.740 | ~SID | 0.775 | 0.768 |
| SVA | 0.830 | 0.832 | SVA | 0.728 | 0.516 |
| ~SVA | 0.517 | 0.728 | ~SVA | 0.763 | 0.761 |

### 6.4.3. Results

Table 10 provides an overview of the findings, using simplified illustrations: "black circles denote the presence of a particular condition (●), crossed-out circles (⊗) signify its absence, and blank spaces represent a "do not care" situation, indicating the potential presence or absence of a causal condition" (Fiss, 2011). The table also includes information on each solution's raw consistency, similar to a regression coefficient, and coverage scores that reflect the magnitude of effects in hypothesis testing. Moreover, the overall coverage of the solution, analogous to the $R^2$ value in variable-based methodologies, helps assess the influence of identified configurations on user acceptance.

The findings reveal two configurations associated with high acceptance and one configuration linked to low acceptance, meeting acceptable consistency (>0.8) and coverage (>0.2) levels, as recommended by Rasoolimanesh et al. (2021).

The first configuration leading to high acceptance represents a group of users who have embraced Rabbit R1 because of its *perceived usefulness × ease of use × conversational intelligence × task intelligence × simplicity in information design × and visual aesthetics.* They do not express any preference for the naturalness of speech (perceived naturalness), meaning that its presence or absence does not influence the configuration. This configuration boasts a consistency score of 0.94 and a raw coverage of 0.521, indicating that 52.1% of respondents chose this combination.

Configuration 2 represents a group of users who have embraced Rabbit R1 because of its *perceived usefulness × conversational intelligence × task intelligence × perceived naturalness × simplicity in information design × and visual aesthetics.* They do not express any preference for the ease of use. This configuration has a consistency score of 0.96 and a raw coverage of 0.537, indicating that 53.7% of respondents chose this combination.

For the configuration leading to low acceptance, the findings reveal that the absence of any factors, regardless of the presence or absence of PEOU, will negatively impact user acceptance of Rabbit R1.

Table 10 Sufficient configurations for high and low user acceptance

| Configurations | Solutions for high ACC | | Solutions for Low ~ACC |
|---|---|---|---|
| | Model: ACC = f (PU, PEU, CI, TI, PN, SID, SVA) Frequency cut-off: 34 Consistency cut-off: 0.85 | | Model: ~ACC= f (PU, PEU, CI, TI, PN, SID, SVA) Frequency cut-off: 34 Consistency cut-off: 0.85 |
| | 1 | 2 | 1 |
| PU | ● | ● | ⊗ |
| PEOU | ● | | |
| CI | ● | ● | ⊗ |
| TI | ● | ● | ⊗ |
| PN | | ● | ⊗ |
| SID | ● | ● | ⊗ |
| SVA | ● | ● | ⊗ |
| Raw coverage | 0.521 | 0.537 | 0.527 |
| Unique coverage | 0.034 | 0.050 | 0.527 |
| Consistency | 0.940 | 0.936 | 0.962 |
| solution coverage | 0.571 | | 0.527 |
| solution consistency | 0.933 | | 0.962 |

## 7. Discussion

This research explores the factors driving user acceptance of PIPAs like Rabbit R1. The study is particularly timely as users are increasingly integrating PIPAs into their daily routines, and stakeholders must understand these critical factors to develop products and services that meet users' needs and expectations. We leverage the TAM framework and extend it with perceived intelligence

factors—conversational intelligence, task intelligence, and perceived naturalness—and simplicity factors—information design and visual aesthetics. The model also includes user loyalty.

Results from PLS-SEM revealed that PU and PEOU significantly and positively impacted user acceptance of PIPAs, indicating that users highly value the practical benefits and efficiency offered by PIPAs. They are more likely to adopt these AI companions when they believe the devices can enhance their productivity and are user-friendly (Balakrishnan & Dwivedi, 2024; Moussawi et al., 2021). These findings align with prior studies that showed that PU and PEOU are critical drivers of IPA adoption intention (Balakrishnan & Dwivedi, 2024; Choi & Kim, 2023; Sohn & Kwon, 2020).

Additionally, conversational intelligence was positively related to PU and PIPA acceptance, suggesting that meaningful and effective conversations with PIPAs enhance their perceived value and overall acceptance. These findings complement those of Ling et al. (2023), which highlighted that conversational intelligence leads to intentions to book and search for information. Task intelligence strongly influenced PU and PIPA acceptance, suggesting that PIPAs' ability to execute tasks proficiently and autonomously is crucial for their perceived value and acceptance. When Rabbit R1 demonstrates high task intelligence, users perceive the device as significantly enhancing their productivity and efficiency (Hu et al., 2021a). This proficiency in task execution reduces the effort and time required for users to complete tasks manually and fosters a sense of reliability and efficiency, contributing to higher acceptance of the PIPA. These findings extend those of Guha et al. (2023), who found that a broader task range increases the perception of intelligence and contributes to positive user evaluation. It also complements the findings of Hu et al. (2021a), who highlighted that IPAs' action autonomy enhances the perception of warmth.

Furthermore, perceived naturalness positively influenced PIPA acceptance, indicating that users are more likely to adopt PIPAs like Rabbit R1 when they have human-like voices. When users perceive the Rabbit R1 as having a natural and human-like voice, it reduces the cognitive effort required to engage with the device (Guha et al., 2023; Humphry & Chesher, 2021), making interactions smoother and more intuitive. This finding extends those of Kang et al. (2023), who found that natural speech increases users' emotional attachment to IPAs. Moreover, simplicity in information design strongly impacted PEOU and PIPA acceptance. When a PIPA is designed with a clear, streamlined interface that eliminates unnecessary steps, users find it easier to navigate and use. The straightforward presentation of information reduces cognitive load and potential frustration (Choi et al., 2012), making the AI companion more user-friendly and suitable for daily use. These findings complement those of Choi et al. (2012), who found that simplicity in information design can lead to user satisfaction.

Moreover, simplicity in visual aesthetics was positively associated with PIPA acceptance, suggesting that a simple and aesthetically pleasing design can significantly enhance user acceptance of PIPAs like Rabbit R1. A visually simple interface, characterized by minimalistic design elements and a coherent visual hierarchy, makes the device more attractive and engaging to users (Choi et al., 2012; Verkijika & De Wet, 2019). This finding aligns with Bölen (2020) and Kumar et al. (2021), who found that visual aesthetics can influence users' intentions. It also extends the findings of Choi et al. (2012) and Verkijika and De Wet (2019), who found that visual aesthetics can lead to user satisfaction.

The findings further highlighted the positive impact of PIPA acceptance on user loyalty, suggesting that users who integrate PIPAs into their daily routines are more likely to develop a sense of attachment and ongoing preference for the AI companions. Loyal customers are more inclined to consistently choose these devices over other alternatives and engage in positive WOM. These findings align with those of Alkhwaldi et al. (2022), who revealed that users who adopt technological products or services often become loyal users. We found that PIPA acceptance mediated the relationship between PU, PEOU, perceived intelligence, simplicity, and user loyalty. This suggests that when users perceive the Rabbit R1 as practical, easy to use, intelligent in conversation and task execution, and simple in design and visual aesthetics, they are more likely to accept the device. This acceptance, in turn, strengthens their loyalty to the product.

Finally, our findings demonstrated the significant role of predictor variables in explaining user acceptance of PIPAs. Through fsQCA analysis, we identified three combinations leading to high and low user acceptance with high coverage and high consistency. Notably, to achieve high acceptance of PIPAs like Rabbit R1, PU, PEOU, conversational intelligence, task intelligence, perceived naturalness, simplicity in information design, and visual aesthetics should be present, as the absence of any of these factors will likely lead to low user acceptance.

### 7.1. Theoretical implications

This research makes several contributions to the IPA and HCI literature. First, most research has traditionally treated perceived intelligence as a single-dimension construct(e.g., Balakrishnan

& Dwivedi, 2024; Moussawi et al., 2021; Priya & Sharma, 2023). In contrast, this study adopts a multidimensional perspective by dissecting perceived intelligence into its underlying factors, including conversational intelligence, task intelligence, and perceived naturalness. This approach allows for a more detailed examination of how these distinct facets of perceived intelligence individually impact user acceptance of IPAs. By analyzing each dimension separately, our research offers deeper insights into the specific aspects of perceived intelligence that contribute to user acceptance and loyalty, thereby advancing the understanding of how intelligent personal assistants are evaluated and accepted in real-world contexts.

This study makes a pioneering contribution by demonstrating the positive impact of conversational intelligence on PU and user acceptance and loyalty with PIPAs. While previous research has shown its influence on users' intentions to book services or search for information (Ling et al., 2023), our study extends this understanding by revealing that conversational intelligence significantly affects users' perceptions of an IPA's usefulness and overall acceptance and loyalty. The findings are also unique in demonstrating the positive influence of task intelligence on PU and user acceptance and loyalty with PIPAs. Unlike previous studies that highlighted the impact of similar constructs (task range and action autonomy) on IPA evaluation and perception of warmth (Guha et al., 2023; Hu et al., 2021a), our research specifically confirms the critical role of task intelligence in enhancing utility perception and driving user acceptance and loyalty with PIPAs. Moreover, we confirmed the impact of perceived naturalness on IPA acceptance and loyalty. While previous research has established its positive influence on emotional attachment to IPAs (Kang et al., 2024), we go further by highlighting that a natural-sounding voice can drive user acceptance and loyalty to IPAs.

This study brings perceived simplicity and its underlying mechanisms—information design and visual aesthetics—to the forefront of IPA research and validates their positive influence on user acceptance and loyalty, highlighting that a simple and minimalistic design is crucial for driving the adoption and long-term engagement with PIPAs. Additionally, while researchers have shown that simplicity in information design can lead to user satisfaction (Choi et al., 2012), this research is the first to demonstrate that simplicity in information design can lead to PEOU, user acceptance, and loyalty. This complements the literature on PEOU, acceptance, and loyalty by adding a new determinant, emphasizing the importance of streamlined, intuitive interfaces in enhancing user experience and promoting the widespread adoption of IPAs.

Finally, our study further advances understanding by demonstrating that user acceptance is a crucial mediator of user loyalty, highlighting its positive mediating role between our antecedent variables and user loyalty. The mediating role of user acceptance has been largely overlooked in prior research. Hence, our study's findings are particularly noteworthy as they offer empirical evidence substantiating this possibility, shedding light on the intricate paths through which PU, PEOU, conversational intelligence, task intelligence, perceived naturalness, simplicity in information design, and visual aesthetics can lead to loyalty. These findings open new avenues for researchers to understand user loyalty in the context of IPAs.

### 7.2. Managerial implications

Based on our findings, we offer several recommendations for system designers and marketers. First, understanding the pivotal role of perceived intelligence attributes, such as conversational intelligence and task intelligence, in driving user acceptance and loyalty suggests that system designers should prioritize these features to ensure a seamless and intelligent interaction with users. This underscores the importance of continuously improving Rabbit R1's ability to engage in meaningful conversations and efficiently execute tasks across diverse applications. Moreover, recognizing the impact of perceived naturalness on user acceptance and loyalty highlights the significance of creating PIPAs with a natural-sounding voice, as it fosters a more relatable and engaging user experience. Marketers of PIPAs should focus on these attributes in their communication strategies to enhance user acceptance and WOM.

Second, given the role of information design and visual aesthetics in influencing PEOU and user acceptance, developers should focus on creating interfaces that are not only aesthetically pleasing but also easy to navigate. This can be achieved by eliminating unnecessary steps in task execution and ensuring a clear visual hierarchy. Third, the demonstrated impact of PIPA acceptance on user loyalty emphasizes the need for sustained engagement and positive user experiences. Companies should invest in continuous improvements and updates to Rabbit R1, ensuring it remains relevant and useful to users over time. Also, fostering a strong emotional connection through personalized interactions and reliable performance can encourage users to become loyal product advocates. Positive WOM generated by satisfied users can significantly enhance the product's market reputation and attract new customers.

Fourth, the mediating role of user acceptance between various antecedent variables (PU, PEOU, perceived intelligence, and perceived simplicity) and user loyalty offers strategic insights for marketers. By focusing on enhancing the factors that drive user acceptance, companies can indirectly boost user loyalty. This comprehensive approach ensures that the product not only meets user needs but also cultivates a loyal customer base, contributing to long-term success in the competitive marketplace. Finally, fsQCA analysis identified two configurations leading to high user acceptance. Therefore, we suggest that system designers and marketers focus on these configurations to increase user acceptance. They should also pay attention to the configuration leading to low user acceptance, as the absence of any contributing factor to high user acceptance will automatically lead to low acceptance.

7.3. Limitations and future research lines

This research has several limitations. First, this research focuses on Rabbit R1, which may limit the broader applicability of the results to other PIPAs with different designs and functionalities. Second, the cross-sectional design employed in this research captures a snapshot of user perceptions and acceptance at a specific time. Future research could conduct a longitudinal study, tracking users from the initial stages of acceptance through to the extended use and experience with the PIPA. Third, we used TAM as the foundational framework. While the extended TAM model includes AI-specific factors and user interface design elements, it does not fully capture the determinants of user acceptance. Future research could explore alternative theoretical frameworks or incorporate additional variables to offer a more comprehensive understanding of user acceptance dynamics.

**Reference**

Acquah, I. S. K., Quaicoe, J., & Gatsi, J. G. (2024). Modelling circular economy capabilities and sustainable manufacturing practices for environmental performance: Assessing linear (PLS-SEM) and non-linear (fsQCA) effects. *Technological Forecasting and Social Change*, *205*, 123501.

Agag, G., Durrani, B. A., Abdelmoety, Z. H., Daher, M. M., & Eid, R. (2024). Understanding the link between net promoter score and e-WOM behaviour on social media: The role of national culture. *Journal of Business Research*, *170*, 114303. https://doi.org/10.1016/j.jbusres.2023.114303

Alkhwaldi, A. F., Alharasis, E. E., Shehadeh, M., Abu-AlSondos, I. A., Oudat, M. S., & Bani Atta, A. A. (2022). Towards an understanding of FinTech users' adoption: Intention and e-loyalty post-COVID-19 from a developing country perspective. *Sustainability*, *14*(19), 12616.

Attié, E., & Meyer-Waarden, L. (2022). The acceptance and usage of smart connected objects according to adoption stages: an enhanced technology acceptance model integrating the diffusion of innovation, uses and gratification and privacy calculus theories. *Technological Forecasting and Social Change*, *176*, 121485. https://doi.org/10.1016/j.techfore.2022.121485

Aw, E. C.-X., Tan, G. W.-H., Cham, T.-H., Raman, R., & Ooi, K.-B. (2022). Alexa, what's on my shopping list? Transforming customer experience with digital voice assistants. *Technological Forecasting and Social Change*, *180*, 121711.

Badghish, S., Shaik, A. S., Sahore, N., Srivastava, S., & Masood, A. (2024). Can transactional use of AI-controlled voice assistants for service delivery pickup pace in the near future? A social learning theory (SLT) perspective. *Technological Forecasting and Social Change*, *198*, 122972.

Balakrishnan, J., & Dwivedi, Y. K. (2024). Conversational commerce: entering the next stage of AI-powered digital assistants. *Annals of Operations Research*, *333*(2), 653–687.

Balakrishnan, J., Abed, S. S., & Jones, P. (2022). The role of meta-UTAUT factors, perceived anthropomorphism, perceived intelligence, and social self-efficacy in chatbot-based services? *Technological Forecasting and Social Change*, *180*, 121692. https://doi.org/10.1016/j.techfore.2022.121692

Bhutoria, A. (2022). Personalized education and artificial intelligence in the United States, China, and India: A systematic review using a human-in-the-loop model. *Computers and Education: Artificial Intelligence*, *3*, 100068. https://doi.org/10.1016/j.caeai.2022.100068

Bölen, M. C. (2020). Exploring the determinants of users' continuance intention in smartwatches. *Technology in Society*, *60*, 101209.

Brenner, W. (2024). Rabbit R1 makes $10M in sales in one week. Available at (https://www.updateordie.com/eng/2024/01/19/rabbit-r1-makes-us10m-in-sales-in-one-week/) (Last accessed august 8, 2024)

Brill, T. M., Munoz, L., & Miller, R. J. (2022). Siri, Alexa, and other digital assistants: a study of customer satisfaction with artificial intelligence applications. In *The role of smart technologies in decision making* (pp. 35–70). Routledge.

Cadenas, C. (2024). Rabbit R1 hasn't shipped yet and it's already received a major upgrade. Available at (https://www.techradar.com/computing/rabbit-r1-hasnt-shipped-yet-and-its-already-received-a-major-upgrade) (Last accessed August 8, 2024)

Camilleri, M. A. (2024). Factors affecting performance expectancy and intentions to use ChatGPT: Using SmartPLS to advance an information technology acceptance framework. *Technological Forecasting and Social Change*, *201*, 123247.

Canbek, N. G., & Mutlu, M. E. (2016). On the track of artificial intelligence: Learning with intelligent personal assistants. *Journal of Human Sciences*, *13*(1), 592–601.

Chen, X., Guo, S., Xiong, J., & Ye, Z. (2023). Customer engagement, dependence and loyalty: An empirical study of Chinese customers in multitouch service encounters. *Technological Forecasting and Social Change*, *197*, 122920.

Choi, J. H., & Lee, H.-J. (2012). Facets of simplicity for the smartphone interface: A structural model. *International Journal of Human-Computer Studies*, *70*(2), 129–142. https://doi.org/10.1016/j.ijhcs.2011.09.002

Choi, S., Jang, Y., & Kim, H. (2023). Exploring factors influencing students' intention to use intelligent personal assistants for learning. *Interactive Learning Environments*, 1–14. https://doi.org/10.1080/10494820.2023.2194927

Cui, Y., Li, J., & Zhang, Y. (2022). The impacts of game experience and fanwork creation on game loyalty: Mediation effect of perceived value. *Technological Forecasting and Social Change*, *176*, 121495.

Davis, F. D. (1989). Perceived Usefulness, Perceived Ease of Use, and User Acceptance of Information Technology. *MIS quarterly,* 13(3), 319-340.

Fiss, P. C. (2011). Building better causal theories: A fuzzy set approach to typologies in organization research. *Academy of Management Journal*, *54*(2), 393–420. https://doi.org/10.5465/amj.2011.60263120

Guha, A., Bressgott, T., Grewal, D., Mahr, D., Wetzels, M., & Schweiger, E. (2023). How artificiality and intelligence affect voice assistant evaluations. *Journal of the Academy of Marketing Science*, *51*(4), 843–866.

Gursoy, D., Chi, O. H., Lu, L., & Nunkoo, R. (2019). Consumers acceptance of artificially intelligent (AI) device use in service delivery. *International Journal of Information Management*, *49*, 157–169.

Hair, J. F., Risher, J. J., Sarstedt, M., & Ringle, C. M. (2019). When to use and how to report the results of PLS-SEM. *European Business Review*, *31*(1), 2–24.

Hu, P., & Lu, Y. (2021). Dual humanness and trust in conversational AI: A person-centered approach. *Computers in Human Behavior*, *119*, 106727.

Hu, Q., Lu, Y., Pan, Z., Gong, Y., & Yang, Z. (2021). Can AI artifacts influence human cognition? The effects of artificial autonomy in intelligent personal assistants. *International Journal of Information Management*, *56*, 102250.

Huang, T.-L., Tsiotsou, R. H., & Liu, B. S. (2023). Delineating the role of mood maintenance in augmenting reality (AR) service experiences: an application in tourism. *Technological Forecasting and Social Change*, *189*, 122385. https://doi.org/10.1016/j.techfore.2023.122385

Humphry, J., & Chesher, C. (2021). Preparing for smart voice assistants: Cultural histories and media innovations. *New Media & Society*, *23*(7), 1971–1988.

Jones, T. O., & Sasser, W. E., Jr. (1995). Why satisfied customers defect. Harvard Business Review, 73(6), 88e99.

Kang, W., Shao, B., Du, S., Chen, H., & Zhang, Y. (2024). How to improve voice assistant evaluations: Understanding the role of attachment with a socio-technical systems perspective. *Technological Forecasting and Social Change*, *200*, 123171. https://doi.org/10.1016/j.techfore.2023.123171

Kautish, P., Purohit, S., Filieri, R., & Dwivedi, Y. K. (2023). Examining the role of consumer motivations to use voice assistants for fashion shopping: The mediating role of awe experience and eWOM. *Technological Forecasting and Social Change*, *190*, 122407. https://doi.org/10.1016/j.techfore.2023.122407

Kaya, F., Aydin, F., Schepman, A., Rodway, P., Yetişensoy, O., & Demir Kaya, M. (2024). The roles of personality traits, AI anxiety, and demographic factors in attitudes toward artificial intelligence. *International Journal of Human–Computer Interaction*, *40*(2), 497–514. https://doi.org/10.1080/10447318.2022.2151730

Kim, J., Merrill Jr, K., & Collins, C. (2021). AI as a friend or assistant: The mediating role of perceived usefulness in social AI vs. functional AI. *Telematics and Informatics*, *64*, 101694. https://doi.org/10.1016/j.tele.2021.101694

Kock, N. (2015). Common method bias in PLS-SEM: A full collinearity assessment approach. *International Journal of E-Collaboration (Ijec)*, *11*(4), 1–10. https//doi.org10.4018/ijec.2015100101

Kopplin, C. S. (2022). Chatbots in the workplace: A technology acceptance study applying uses and gratifications in coworking spaces. *Journal of Organizational Computing and Electronic Commerce*, *32*(3–4), 232–257.

Kumar, S., Jain, A., & Hsieh, J.-K. (2021). Impact of apps aesthetics on revisit intentions of food delivery apps: The mediating role of pleasure and arousal. *Journal of Retailing and Consumer Services*, *63*, 102686.

Lew, S., Tan, G. W.-H., Loh, X.-M., Hew, J.-J., & Ooi, K.-B. (2020). The disruptive mobile wallet in the hospitality industry: An extended mobile technology acceptance model. *Technology in Society*, *63*, 101430. https://doi.org/10.1016/j.techsoc.2020.101430

Liao, Y., Vitak, J., Kumar, P., Zimmer, M., & Kritikos, K. (2019). Understanding the role of privacy and trust in intelligent personal assistant adoption. *Information in Contemporary Society: 14th International Conference, IConference 2019, Washington, DC, USA, March 31–April 3, 2019, Proceedings 14*, 102–113.

Ling, E. C., Tussyadiah, I., Liu, A., & Stienmetz, J. (2023). Perceived Intelligence of Artificially Intelligent Assistants for Travel: Scale Development and Validation. *Journal of Travel Research*, 00472875231217899. https://doi.org/10.1177/00472875231217899

Ling, E. C., Tussyadiah, I., Tuomi, A., Stienmetz, J., & Ioannou, A. (2021). Factors influencing users' adoption and use of conversational agents: A systematic review. *Psychology & Marketing*, *38*(7), 1031–1051. https://doi.org/10.1002/mar.21491

Luo, L., Liu, L., Zheng, Y., & Chen, J. (2024). Visual information and appearance: The impact of visual attributes of user-generated photos on review helpfulness. *Telematics and Informatics*, 102164.

Maduku, D. K., Mpinganjira, M., Rana, N. P., Thusi, P., Ledikwe, A., & Mkhize, N. H. (2023). Assessing customer passion, commitment, and word-of-mouth intentions in digital assistant usage: The moderating role of technology anxiety. *Journal of Retailing and Consumer Services*, *71*, 103208. https://doi.org/10.1016/j.jretconser.2022.103208

Maroufkhani, P., Asadi, S., Ghobakhloo, M., Jannesari, M. T., & Ismail, W. K. W. (2022). How do interactive voice assistants build brands' loyalty? *Technological Forecasting and Social Change*, *183*, 121870.

Martin, W. C., & Lueg, J. E. (2013). Modeling word-of-mouth usage. *Journal of Business Research*, *66*(7), 801–808. https://doi.org/10.1016/j.jbusres.2011.06.004

McLean, G., & Osei-Frimpong, K. (2019). Hey Alexa… examine the variables influencing the use of artificial intelligent in-home voice assistants. *Computers in Human Behavior*, *99*, 28–37.

Medium. (2024) What is Rabbit OS and Why is it Generating So Much Buzz? Available at: (https://medium.com/@hustle-smart-with-technology/what-is-rabbit-os-and-why-is-it-generating-so-much-buzz-141c4973f374) (Last accessed August 8, 2024)

Moussawi, S., & Koufaris, M. (2019). *Perceived intelligence and perceived anthropomorphism of personal intelligent agents: Scale development and validation*. http://hdl.handle.net/10125/59452

Moussawi, S., Koufaris, M., & Benbunan-Fich, R. (2021). How perceptions of intelligence and anthropomorphism affect adoption of personal intelligent agents. *Electronic Markets*, *31*, 343–364. https://doi.org/10.1007/s12525-020-00411-w

Mvondo, G. F. N, Jing, F., Hussain, K., & Raza, M. A. (2022b). Converting tourists into evangelists: Exploring the role of tourists' participation in value co-creation in enhancing brand evangelism, empowerment, and commitment. *Journal of Hospitality and Tourism Management*, *52*, 1–12. https://doi.org/10.1016/j.jhtm.2022.05.015

Mvondo, G. F. N., Jing, F., Hussain, K., Jin, S., & Raza, M. A. (2022). Impact of international tourists' co-creation experience on brand trust, brand passion, and brand evangelism. *Frontiers in Psychology*, *13*.13:866362. https://doi.org/10.3389/fpsyg.2022.866362

Mvondo, G. F. N., & Niu, B. (2024). Factors Influencing User Willingness To Use SORA. *ArXiv Preprint ArXiv:2405.03986*.

Mvondo, G. F. N., Niu, B., & Eivazinezhad, S. (2023). Exploring The Ethical Use of LLM Chatbots In Higher Education. *Available at SSRN4548263 (2023)*.

Navarro-García, A., Ledesma-Chaves, P., Gil-Cordero, E., & De-Juan-Vigaray, M. D. (2024). Intangible resources, static and dynamic capabilities and perceived competitive advantage in exporting firms. A PLS-SEM/fsQCA approach. *Technological Forecasting and Social Change*, *198*, 123001.

Niu, B., & Mvondo, G. F. N. (2024). I Am ChatGPT, the ultimate AI Chatbot! Investigating the determinants of users' loyalty and ethical usage concerns of ChatGPT. *Journal of Retailing and Consumer Services*, *76*, 103562. https://doi.org/10.1016/j.jretconser.2023.103562

Okine, A. N. D., Li, Y., Djimesah, I. E., Zhao, H., Budu, K. W. A., Duah, E., & Mireku, K. K. (2023). Analyzing crowdfunding adoption from a technology acceptance perspective. *Technological Forecasting and Social Change*, *192*, 122582. https://doi.org/10.1016/j.techfore.2023.122582

Pang, H., Ruan, Y., & Zhang, K. (2024). Deciphering technological contributions of visibility and interactivity to website atmospheric and customer stickiness in AI-driven websites: The pivotal function of online flow state. *Journal of Retailing and Consumer Services*, *78*, 103795.

Pappas, I. O., Kourouthanassis, P. E., Giannakos, M. N., & Chrissikopoulos, V. (2016). Explaining online shopping behavior with fsQCA: The role of cognitive and affective perceptions. *Journal of Business Research*, *69*(2), 794–803. https://doi.org/10.1016/j.jbusres.2015.07.010

Priya, B., & Sharma, V. (2023). Exploring users' adoption intentions of intelligent virtual assistants in financial services: An anthropomorphic perspectives and socio-psychological perspectives. *Computers in Human Behavior*, *148*, 107912. https://doi.org/10.1016/j.chb.2023.107912

Ragin, C. C. (2009). *Redesigning social inquiry: Fuzzy sets and beyond*. University of Chicago Press.

Rasoolimanesh, S. M., Ringle, C. M., Sarstedt, M., & Olya, H. (2021). The combined use of symmetric and asymmetric approaches: Partial least squares-structural equation modeling and fuzzy-set qualitative comparative analysis. *International Journal of Contemporary*

*Hospitality Management*, *33*(5), 1571–1592. https://doi.org/10.1108/IJCHM-10-2020-1164

Rodríguez-López, M. E., Higueras-Castillo, E., Rojas-Lamorena, Á. J., & Alcántara-Pilar, J. M. (2024). The future of TV-shopping: predicting user purchase intention through an extended technology acceptance model. *Technological Forecasting and Social Change*, *198*, 122986. https://doi.org/10.1016/j.techfore.2023.122986

Santiago, J., Borges-Tiago, M. T., & Tiago, F. (2024). Embracing RAISA in restaurants: Exploring customer attitudes toward robot adoption. *Technological Forecasting and Social Change*, *199*, 123047. https://doi.org/10.1016/j.techfore.2023.123047

Sharma, M., Joshi, S., Luthra, S., & Kumar, A. (2024). Impact of digital assistant attributes on millennials' purchasing intentions: A multi-group analysis using PLS-SEM, artificial neural network and fsQCA. *Information Systems Frontiers*, *26*(3), 943–966.

Sohaib, M., & Han, H. (2023). Building value co-creation with social media marketing, brand trust, and brand loyalty. *Journal of Retailing and Consumer Services*, *74*, 103442.

Sohn, K., & Kwon, O. (2020). Technology acceptance theories and factors influencing artificial Intelligence-based intelligent products. *Telematics and Informatics*, *47*, 101324. https://doi.org/10.1016/j.tele.2019.101324

Song, M., Xing, X., Duan, Y., Cohen, J., & Mou, J. (2022). Will artificial intelligence replace human customer service? The impact of communication quality and privacy risks on adoption intention. *Journal of Retailing and Consumer Services*, *66*, 102900. https://doi.org/10.1016/j.jretconser.2021.102900

Sullivan, M. (2024). "Rabbit's handled device aims to create a post-smartphone experiene." Available at: (https://www.fastcompany.com/91007453/rabbits-handheld-ai-device-aims-to-create-a-post-smartphone-experience) (Last accessed August 8, 2024)

Sun, C.-C. (2021). Analyzing determinants for adoption of intelligent personal assistant: An empirical study. *Applied Sciences*, *11*(22), 10618. https://doi.org/10.3390/app112210618

The verge. (2024). The Rabbit R1 is an AI-powered gadget that can use your apps for you. Available at: (https://www.theverge.com/2024/1/9/24030667/rabbit-r1-ai-action-model-price-release-date) (Last accessed August 8, 2024).

Türker, C., Altay, B. C., & Okumuş, A. (2022). Understanding user acceptance of QR code mobile payment systems in Turkey: An extended TAM. *Technological Forecasting and Social Change*, *184*, 121968.

Verkijika, S. F., & De Wet, L. (2019). Understanding word-of-mouth (WOM) intentions of mobile app users: The role of simplicity and emotions during the first interaction. *Telematics and Informatics*, *41*, 218–228. https://doi.org/10.1016/j.tele.2019.05.003

Vimalkumar, M., Sharma, S. K., Singh, J. B., & Dwivedi, Y. K. (2021). 'Okay google, what about my privacy?': User's privacy perceptions and acceptance of voice based digital assistants. *Computers in Human Behavior*, *120*, 106763.

Waytz, A., Morewedge, C. K., Epley, N., Monteleone, G., Gao, J.-H., & Cacioppo, J. T. (2010). Making sense by making sentient: effectance motivation increases anthropomorphism. *Journal of Personality and Social Psychology*, *99*(3), 410. https://doi.org/10.1037/a0020240

Xie, C., Wang, Y., & Cheng, Y. (2024). Does artificial intelligence satisfy you? A meta-analysis of user gratification and user satisfaction with AI-powered chatbots. *International Journal of Human–Computer Interaction*, *40*(3), 613–623.

Zhao, Y., Chen, Y., Sun, Y., & Shen, X.-L. (2024). Understanding users' voice assistant exploration intention: unraveling the differential mechanisms of the multiple dimensions of perceived intelligence. *Internet Research*. https://doi.org/10.1108/INTR-10-2022-0807

Zulhusni, M. (2024). What is exactly the rabbit R1 AI device that confused many people? Available at: (https://techwireasia.com/01/2024/rabbit-r1-the-future-of-ai-assisted-living-starts-here/) (Last accessed August 8, 2024)

Zou, Y., Zhang, X., Zheng, W., & Huang, Z. (2023). Exploring the sustainable influencing factors of audience loyalty of Chinese sports live broadcast platform based on SEM model. *Technological Forecasting and Social Change*, *189*, 122362.